\documentclass[12pt]{article}
\usepackage{a4wide}
\usepackage{amssymb}
\usepackage{amsmath}
\usepackage{graphicx}
\usepackage{mathdots}
\usepackage{slashed}
\usepackage{verbatim}
\usepackage{xcolor}

\usepackage{hyperref}

\usepackage{IEEEtrantools}

\def\makeatletter{\catcode`\@=11}
\makeatletter
\def\mathbox#1{\hbox{$\m@th#1$}}%
\def\math@ccstyles#1#2#3#4#5#6#7{{\leavevmode
      \setbox0\mathbox{#6#7}%
      \setbox2\mathbox{#4#5}%
      \dimen@ #3%
      \baselineskip\z@\lineskiplimit#1\lineskip\z@
      \vbox{\ialign{##\crcr
             \hfil \kern #2\box2 \hfil\crcr
             \noalign{\kern\dimen@}%
             \hfil\box0\hfil\crcr}}}}
\def\mathaccstyles{\math@ccstyles\maxdimen}
\def\maththroughstyles{\math@ccstyles{-\maxdimen}}
\def\unity%
 {\maththroughstyles{.45\ht0}\z@\displaystyle {\mathchar"006C}\displaystyle 1}
 
\begin{document}

\setcounter{table}{0}

\mbox{}
\vspace{2truecm}
\linespread{1.1}

\centerline{\LARGE \bf A note on instanton operators,}

\vspace{.5cm}

 \centerline{\LARGE \bf instanton particles, and supersymmetry}

\vspace{2truecm}

\centerline{
    {\large \bf Oren Bergman ${}^{a}$} \footnote{bergman@physics.technion.ac.il}
     {\bf and}
    {\large \bf Diego Rodriguez-Gomez${}^{b}$} \footnote{d.rodriguez.gomez@uniovi.es}}

\vspace{1cm}
\centerline{{\it ${}^a$ Department of Physics, Technion, Israel Institute of Technology}} \centerline{{\it Haifa, 32000, Israel}}
\vspace{1cm}
\centerline{{\it ${}^b$ Department of Physics, Universidad de Oviedo}} \centerline{{\it Avda.~Calvo Sotelo 18, 33007}} \centerline{{\it Oviedo, Spain}} 
\vspace{1cm}

\centerline{\bf ABSTRACT}
\vspace{1truecm}

\noindent We clarify certain aspects of instanton operators in five-dimensional supersymmetric gauge theories. In particular, we show how, in the pointlike limit, they become supersymmetric and provide the natural bridge with the instantonic states contributing to the index as well as with the zero mode counting leading to broken current multiplets.

\newpage

\section{Introduction and summary}

An important feature of gauge theories in five dimensions is that they come equipped with a topologically conserved current
(one for each simple gauge group factor)
 
\begin{equation}
\label{instanton_current}
J=\frac{1}{8\pi^2} \star {\rm Tr} F\wedge F\, .
\end{equation}
The associated conserved charge is carried by solitonic states that correspond to the lift of BPST instantons in $\mathbb{R}^4$.
As such, these states are expected to have a mass of ${\cal O}(1/g^2)$, and therefore to become important for low energy physics at strong coupling.
Alternatively we can consider a local operator in $\mathbb{R}^5$ such that 
\begin{equation}
\label{instanton_flux}
\frac{1}{8\pi^2} \int_{S^4} {\rm Tr} F\wedge F = n\,,
\end{equation}
where the $S^4$ surrounds the location of the operator.
This is called an $n$-instanton operator.
These two objects, corresponding to the gauge field configurations on $\mathbb{R}^4$ and $S^4$, 
can be related by a conformal map \cite{Constable:2001ag}.
Note that this is not the usual state-operator map, which would relate the local operator in ${\mathbb R}^5$ to a state 
in $S^4\times \mathbb{R}$, namely in the radially quantized theory.

Gauge theories in five dimensions are non-renormalizable and therefore do not make sense, at least in perturbation theory,
as microscopic quantum field theories. In general this is an indication that the theory is not UV complete, and requires
the addition of extra degrees of freedom.
However it has become apparent, using both supersymmetric dynamics and string theory,
that supersymmetric gauge theories in five dimensions can in some cases be UV complete,
and possess a UV fixed point corresponding to a superconformal field theory (SCFT) in either five dimensions \cite{Seiberg:1996bd}
or six dimensions \cite{Douglas:2010iu}. Indeed, we may think of the set of microscopically well-defined theories in five or six dimensions as 
the IR theories resulting from relevant deformations of these fixed point theories.\footnote{Superconformal theories in five and six dimensions
have no marginal deformations, and only relevant deformations are available.}
One particularly interesting deformation corresponds to turning on a Yang-Mills coupling; a mass deformation triggering a flow to an IR 
gauge theory.

In many cases instantons play an important role in identifying the fixed point theory, and in particular its symmetries.
For example, the UV fixed point of the maximally supersymmetric 5d Yang-Mills theory is conjectured to
be the 6d $(2,0)$ SCFT compactified on a circle in the limit of infinite radius.
It has been argued that multi-instanton bound states in the 5d gauge theory reproduce the entire KK spectrum
of the 6d theory \cite{Lambert:2010iw}. The corresponding instanton operators should therefore incorporate
the currents associated to the broken 6d spacetime symmetry.
Further support for this comes from analyzing the zero-modes of the 5d instanton \cite{Tachikawa:2015mha}. A somewhat similar phenomenon occurs in cases where the UV fixed point is a 5d SCFT, where the internal global
symmetry is enhanced in some examples by instantons.
In some of these examples the enhancement can be explicitly shown by computing the superconformal index \cite{Kim:2012gu},
and by a zero-mode analysis \cite{Tachikawa:2015mha,Zafrir:2015uaa,Yonekura:2015ksa,Gaiotto:2015una} 
(see also \cite{Cremonesi:2015lsa} for an alternative approach).

Inherent to this discussion is the requirement that the instanton states and operators come in
short (BPS) supersymmetry multiplets, so that their masses and dimensions are protected.
This is indeed true for the solitonic state obtained by lifting the BPST configuration to five dimensions. On the other hand it has recently been pointed out that instanton operators do not preserve any supersymmetry,
and therefore belong to long multiplets of operators \cite{Lambert:2014jna} (see also \cite{Rodriguez-Gomez:2015xwa}).
This also appears to be consistent with the counting of fermionic zero modes associated to the spacetime symmetry \cite{Tachikawa:2015mha}.
But it is clearly in conflict with the computations of 5d superconformal indices that include crucial contributions of instanton operators.

The purpose of this note is to resolve this apparent contradiction.\footnote{Our resolution is briefly mentioned
in a revised version of \cite{Lambert:2014jna}.}
A key observation is that there is a one parameter family of instanton operators,
labelled by a dimensionless parameter $\rho$ that determines how the flux is distributed on $S^4$.
Generically the configuration breaks all the supersymmetry, however in the limit $\rho\rightarrow 0$ (or $\rho\rightarrow \infty$)
half of the supersymmetry is preserved.
In this limit the flux is concentrated at the north or south pole of $S^4$, depending on the sign of $n$.
This makes contact with the supersymmetric property of the instanton state, since
the state created by the operator in this limit is the pointlike limit of the BPST state in the $\mathbb{R}^4$ tangent to the pole.
It also makes contact with the superconformal index computations, where it was shown that the instanton (anti-instanton) configurations
contributing to the index are necessarily concentrated at the south (north) pole of $S^4$. Moreover, by regarding the mass deformation to the Yang-Mills theory as spontaneous symmetry breaking, the structure of zero modes necessary to find the expected broken current multiplet naturally follows.

In the rest of this note we will offer support for our claim. 
For simplicity we focus on the ${\cal N}=1$ $SU(2)$ theory.

\section{Instanton operators}

Consider a local operator  
that inserts $n$ units of instanton-flux on an $S^4$ enclosing the point $x=0$ in $\mathbb{R}^5$, Eq.~(\ref{instanton_flux}). A classical configuration satisfying this with $n=1$ can easily be constructed starting with the $SU(2)$ BPST solution in ${\mathbb R}^4$
and conformally mapping it to $S^4$.
The BPST solution is given by
\begin{equation}
F_{ij} = \frac{2\mu^2 \epsilon_{ijk} \sigma_k}{\left(|{x}|^2 + \mu^2\right)^2} \quad , \quad
F_{i4} = \frac{2\mu^2 \sigma_i}{\left(|{x}|^2 + \mu^2\right)^2} 
\end{equation}
where $\mu$ parameterizes the size of the instanton.
This satisfies 
\begin{equation}
\label{BPST}
\left.{\rm Tr}(F\wedge F)\right|_{{\mathbb R}^4}= \mbox{} - 96\,\frac{\mu^4}{(|{x}|^2+\mu^2)^4} d^4 x\,.
\end{equation}
Conformally mapping the configuration to an $S^4$ with metric

\begin{equation}
d\Omega_4^2=R^2\,\Big[d\alpha_1^2+\sin^2\alpha_1(d\alpha_2^2+\sin^2\alpha_2(d\alpha_3^2+\sin^2\alpha_3 \,d\alpha^2_4)\Big)\Big]\, ,
\end{equation}
gives
\begin{equation}
\label{F2polar}
\left.{\rm Tr}(F\wedge F)\right|_{S^4}=96\,\frac{\rho^4}{\Big((1+\rho^2)+(1-\rho^2)\,\cos\alpha_1\Big)^4}\, \omega_4\,,
\end{equation}
where $\omega_4$ is the volume form of the unit $S^4$, and $\rho = \mu /R$.
The dimensionless parameter $\rho$ determines how the flux is distributed on $S^4$.
In particular $\rho =1$ corresponds to a uniform distribution, and in the limits $\rho\rightarrow 0$ and $\infty$
the flux is concentrated at $\alpha_1=\pi$ and at $\alpha_1 = 0$, respectively.

Let us now analyze the supersymmetry of this configuration. The supersymmetry variation of the gaugino  is given by (we will follow the conventions in \cite{Bergshoeff:2004kh}) 
\begin{equation}
\label{SUSY}
 \delta\Omega = \mbox{} -\slashed{F}\,(\unity\mp \Gamma_r)\epsilon\,.
\end{equation}
We use polar coordinates, in which the 5d instanton field strength satisfies
$F_{ri}=0$ and $F|_{S^4}=\pm \star_{4} F|_{S^4}$, and $\epsilon$ is a 5d background Killing spinor. 
The vacuum of the 5d SCFT has two sets of Killing spinors associated to the Poincar\'e and superconformal supersymmetries.
In polar coordinates these are given  respectivley by\footnote{The superconformal Killing spinors
arise as solutions to five-dimensional conformal supergravity \cite{Kuzenko:2014eqa,Alday:2015lta,Pini:2015xha}}
\begin{eqnarray}
\label{spinorsinpolar}
\epsilon_q &=& e^{\frac{\alpha_1}{2}\,\Gamma_{r1}}\,e^{\frac{\alpha_2}{2}\,\Gamma_{12}}\,e^{\frac{\alpha_3}{2}\,\Gamma_{23}}\,e^{\frac{\alpha_4}{2}\,\Gamma_{34}}\,\epsilon_{q,0} \\
 \epsilon_s &=& r\,\Gamma_r\,e^{\frac{\alpha_1}{2}\,\Gamma_{r1}}\,e^{\frac{\alpha_2}{2}\,\Gamma_{12}}\,e^{\frac{\alpha_3}{2}\,\Gamma_{23}}\,e^{\frac{\alpha_4}{2}\,\Gamma_{34}}\,\epsilon_{s,0} \,,
\end{eqnarray}
where $\epsilon_{q,0}$ and $\epsilon_{s,0}$ are constant spinors.\footnote{The 5d supercharge $\epsilon$ transforms as a ${\bf 4}$
under $Spin(5) = Sp(4)$ and as a ${\bf 2}$ under $SU(2)_R$. However it satisfies 
a symplectic Majorana condition 
$(\epsilon^i)^{\star}=\epsilon^{ij}C\epsilon^j$, where $C$ the charge conjugation matrix, so there are 8 real supersymmetries.} The supercharges preserved by the instanton configuration will be those for which $\Gamma_r\epsilon=\pm \epsilon$. Clearly there is no solution to this 
everywhere on the $S^4$.
Note however that $\Gamma_r\epsilon = +\epsilon$ at $\alpha_1=0$, and $\Gamma_r\epsilon = - \epsilon$ at $\alpha_1=\pi$. Therefore all supersymmetries are broken by this configuration. 

We can ask what state does this operator create in temporal quantization.
Let us take the ``time" direction to be along the polar axis of the $S^4$, specifically $t= \mbox{} - r\cos\alpha_1$, where the future is in the direction of $\alpha_1=\pi$, and
$r^2=t^2+\vec{x}^2$. 
Note that this corresponds to the natural choice $x^1=t$ in $\mathbb{R}^5$ in cartesian coordinates. Projecting onto a constant future time slice we find 
\begin{equation}
\label{F2cartesian}
\left. {\rm Tr}(F\wedge F)\right|_{{\mathbb R}^4} = \mbox{} -96\,\frac{\mu_{eff}^4}{(\vec{x}^2+\mu_{eff}^2)^4}\, d^4\vec{x} 
\quad \mbox{with} \quad \mu_{eff}=\rho\,(\sqrt{t^2+\vec{x}^2} + t) \,,
\end{equation}
namely a BPST-like configuration with a spacetime dependent size parameter $\mu_{eff}$. From this point of view it is natural that the configuration breaks supersymmetry.

\section{Pointlike instantons}

There are two exceptions to the above result,  $\rho=0$ and $\rho\rightarrow\infty$.
In these limits the flux is concentrated at a single point on the $S^4$, at $\alpha_1=\pi$ for $\rho=0$
and at $\alpha_1=0$ for $\rho\rightarrow\infty$.
Therefore one only needs to satisfy the chirality condition locally at these points.
As noted above, the Killing spinors satisfy $\Gamma_r\epsilon = \epsilon$ at $\alpha_1=0$ and
$\Gamma_r\epsilon = -\epsilon$ at $\alpha_1=\pi$. 
Therefore the instanton operator with $F= \star_{S^4} F$ is supersymmetric for $\rho\rightarrow\infty$,
and the instanton operator with $F= - \star_{S^4} F$ is supersymmetric for $\rho=0$.
This is consistent with what was found in the supersymmetric localization procedure for computing the superconformal index \cite{Kim:2012gu}.
It was shown there that instanton operators contribute to the index, but only if the flux is concentrated at one pole for
self-dual configurations, and at the opposite pole for anti-self-dual configurations.
The corresponding states in temporal quantization are point-like supersymmetric BPST configurations.

More generally if we consider an operator corresponding to a distribution in $\rho$, say with $F= - \star_{S^4} F$, 
then at very late time the $\rho >0$ components will spread out and dissolve, and
only the point-like supersymmetric $\rho=0$ state will remain.
This is in line with the point made in \cite{Lambert:2014jna}, that the BPS state can be extracted from the action of the instanton operator
on the vacuum in the late time limit as
\begin{equation}
|n\rangle = \lim_{\tau\rightarrow\infty} e^{-(H-Q)\tau} {\cal I}_n(0)|0\rangle \,.
\end{equation}

\section{Fermionic zero modes}

Fermionic fields in the presence of instantons have zero modes which determine the global and gauge
symmetry representation of the instanton state.
In particular the gaugino zero modes determine the supermultiplet structure.
These arise generically from the broken supersymmetries in (\ref{SUSY}).
A BPS state, the point-like instanton in our case, preserves a fraction (half in this case) of the supersymmetry and therefore transforms in a short supermultiplet.
For the pure $SU(2)$ theory this gives a broken current multiplet, consisting of an $SO(5)$ vector $J_\mu$,
two spinors $\psi_i$ transforming as a doublet of $SU(2)_R$, and three scalars $\mu_{(ij)}$ in the triplet of $SU(2)_R$ \cite{Tachikawa:2015mha}.
This supermultiplet is generated by eight fermionic zero modes corresponding to four broken Poincar\'e supersymmetries, $\epsilon_q$
with $\Gamma_r\epsilon_q = \mp \epsilon_q$,
and four broken superconformal supersymmetries, $\epsilon_s$ with $\Gamma_r\epsilon_s = \mp \epsilon_s$.
We can demonstrate this explicitly as follows.
In the pointlike limit we can approximate the $S^4$ instanton configuration created by the instanton operator as 
an instanton configuration in the tangent $\mathbb{R}^4$.
In this case the background Killing spinors are given by
\begin{equation}
\epsilon_q = \epsilon_{q,0} \;, \quad \epsilon_s = \slashed{x}\epsilon_{s,0} \,.
\end{equation}
The fermionic zero modes of the instanton configuration should be given by the broken supersymmetries as
\begin{equation}
\label{FZM1}
\delta\Omega_q = \slashed{F} \epsilon_q = \slashed{F} \epsilon_{q,0} \,, \quad \Gamma_5 \epsilon_{q,0} = \mp\epsilon_{q,0} \,,
\end{equation}
and 
\begin{equation}
\label{FZM2}
\delta\Omega_s = \slashed{F} \epsilon_s = \slashed{F} \slashed{x} \epsilon_{s,0}\,, \quad \Gamma_5 \epsilon_{s,0} = \mp\epsilon_{s,0} \,.
\end{equation}
One can then easily check that these two spinors satisfy the massless Dirac equation.

This appears somewhat strange from the point of view of the IR gauge theory, since the entire superconformal symmetry
is explicitly broken by the Yang-Mills action. Nevertheless, the conditions on the fermionic zero modes (\ref{FZM1}) and (\ref{FZM2}) continue to hold in the IR gauge theory.
This is understood from the fact that the
the mass deformation leading to the gauge theory can be thought of as the VEV of a scalar in a background
vector multiplet associated to the topological $U(1)$ symmetry. 
In this sense the superconformal symmetry is broken spontaneously.
Since the gauginos are neutral under this symmetry the VEV does not affect their Dirac equation and therefore 
does not affect the analysis of fermionic zero modes.

An interesting complementary point of view can be obtained by performing the state-operator conformal transformation from $\mathbb{R}^5$
to $S^4\times \mathbb{R}$.
Regarding the YM coupling as the VEV of a scalar field in a background vector multiplet, it should map as $g_0 \rightarrow g_0 e^{-\tau/2}$,
where $\tau$ is the coordinate along $\mathbb{R}$.
This is consistent with the observation made in \cite{Pini:2015xha}, that in $S^4 \times \mathbb{R}$ only a position-dependent gauge coupling ``constant" $g_{YM}=g_0 e^{-\tau/2}$ can be turned on in a supersymmetric way. This is a position-space version of the RG flow interpolating between the UV SCFT at early time and the IR supersymmetric gauge theory at late time. 
Moreover, as shown in \cite{Pini:2015xha}, the supersymmetries preserved by this position-dependent YM coupling are the same as in the fixed point theory. Indeed, this would be the natural way to regularize the index computation of \cite{Kim:2012gu}.

\section*{Acknowledgements}

We would like to thank N.~Lambert and C.~Papageorgakis for useful comments and discussions.
D.R-G. is partly supported by the spanish grant MINECO-13-FPA2012-35043-C02-02, the Ramon y Cajal grant RYC-2011-07593 as well as the EU CIG grant UE-14-GT5LD2013-618459. 
O.B. is supported in part by the Israel Science Foundation under grant no. 352/13,
the German-Israeli Foundation for Scientific Research and Development under grant no.~1156-124.7/2011,
and the US-Israel Binational Science Foundation under grant no. 2012-041.

\end{document}